\documentclass[
reprint,
superscriptaddress,
amsmath,
amssymb,
aps,
prl
]{revtex4-1}

\usepackage{graphicx}
\usepackage{xcolor} 

\begin{document}

\title{Rotational Coherence Times of Polar Molecules in Optical Tweezers}

\author{Sean Burchesky}
\author{Lo\"ic Anderegg}
\author{Yicheng Bao}
\author{Scarlett S. Yu}
\affiliation{Department of Physics, Harvard University, Cambridge, MA 02138, USA}
\affiliation{Harvard-MIT Center for Ultracold Atoms, Cambridge, MA 02138, USA}

\author{Eunmi Chae}
\affiliation{Department of Physics, Korea University, Seongbuk-gu, Seoul, South Korea}

\author{Wolfgang Ketterle}
\affiliation{Harvard-MIT Center for Ultracold Atoms, Cambridge, MA 02138, USA}
\affiliation{Department of Physics, Massachusetts Institute of Technology, Cambridge, MA 02139, USA }

\author{Kang-Kuen Ni} 
\affiliation{Department of Physics, Harvard University, Cambridge, MA 02138, USA}
\affiliation{Harvard-MIT Center for Ultracold Atoms, Cambridge, MA 02138, USA}
\affiliation{Department of Chemistry and Chemical Biology, Harvard University, Cambridge, MA 02138, USA}

\author{John M. Doyle} 
\affiliation{Department of Physics, Harvard University, Cambridge, MA 02138, USA}
\affiliation{Harvard-MIT Center for Ultracold Atoms, Cambridge, MA 02138, USA}

\date{\today}
\begin{abstract}

Qubit coherence times are critical to the performance of any robust quantum computing platform. For quantum information processing using arrays of polar molecules, a key performance parameter is the molecular rotational coherence time. We report a 93(7) ms coherence time for rotational state qubits of laser cooled CaF molecules in optical tweezer traps, over an order of magnitude longer than previous systems. Inhomogeneous broadening due to the differential polarizability between the qubit states is suppressed by tuning the tweezer polarization and applied magnetic field to a ``magic'' angle. The coherence time is limited by the residual differential polarizability, implying improvement with further cooling. A single spin-echo pulse is able to extend the coherence time to nearly half a second. The measured coherence times demonstrate the potential of polar molecules as high fidelity qubits.

\end{abstract}

\maketitle

Ultracold polar molecules can provide a powerful and versatile platform for quantum simulation, quantum computation and precision measurement~\cite{demille02qi,Bohn2017,Carr2009review,Safronova2018}. Advances in the direct laser cooling of molecules~\cite{Shuman2009, barry14, truppe17,anderegg17, collopy18}, assembly of molecules from ultracold atoms~\cite{Ni2008KRb, Takekoshi2014RbCs, Park2015NaK, molony2014, dajun2016, Rvachov2017LiNa, cairncross2021, Ye2018NaRb}, and single state control~\cite{Ospelkaus2010hfcontrol, Park2017NaK, Seeselberg2018rotcoh, Chou2017}, demonstrates a handle over the complicated internal structure of molecules. While the large number of internal states in a molecule can present challenges for laser cooling, the additional rotational and vibrational structure provides distinct advantages for quantum simulation and computation applications~\cite{demille02qi,Ni2018,eunmimgf,Sawant_2020}. For example, microwave addressable rotational levels facilitate robust single-qubit operations and electric dipole coupling between adjacent molecules provides gate operations with a predicted fidelity greater than 99.99$\%$ \cite{Ni2018,Hughes2020}. In addition, nuclear spin states in the ground rotational manifold can safeguard quantum information, acting as viable storage qubits~\cite{Park2017NaK,gregory2021robust}. The combination of long-lived rotational states with strong, switchable, dipolar interactions and non-interacting storage states for long quantum memories render ultracold polar molecules a very appealing qubit platform in a realistic system.

\begin{figure}[b]
\centering
\includegraphics[width=.5\textwidth]{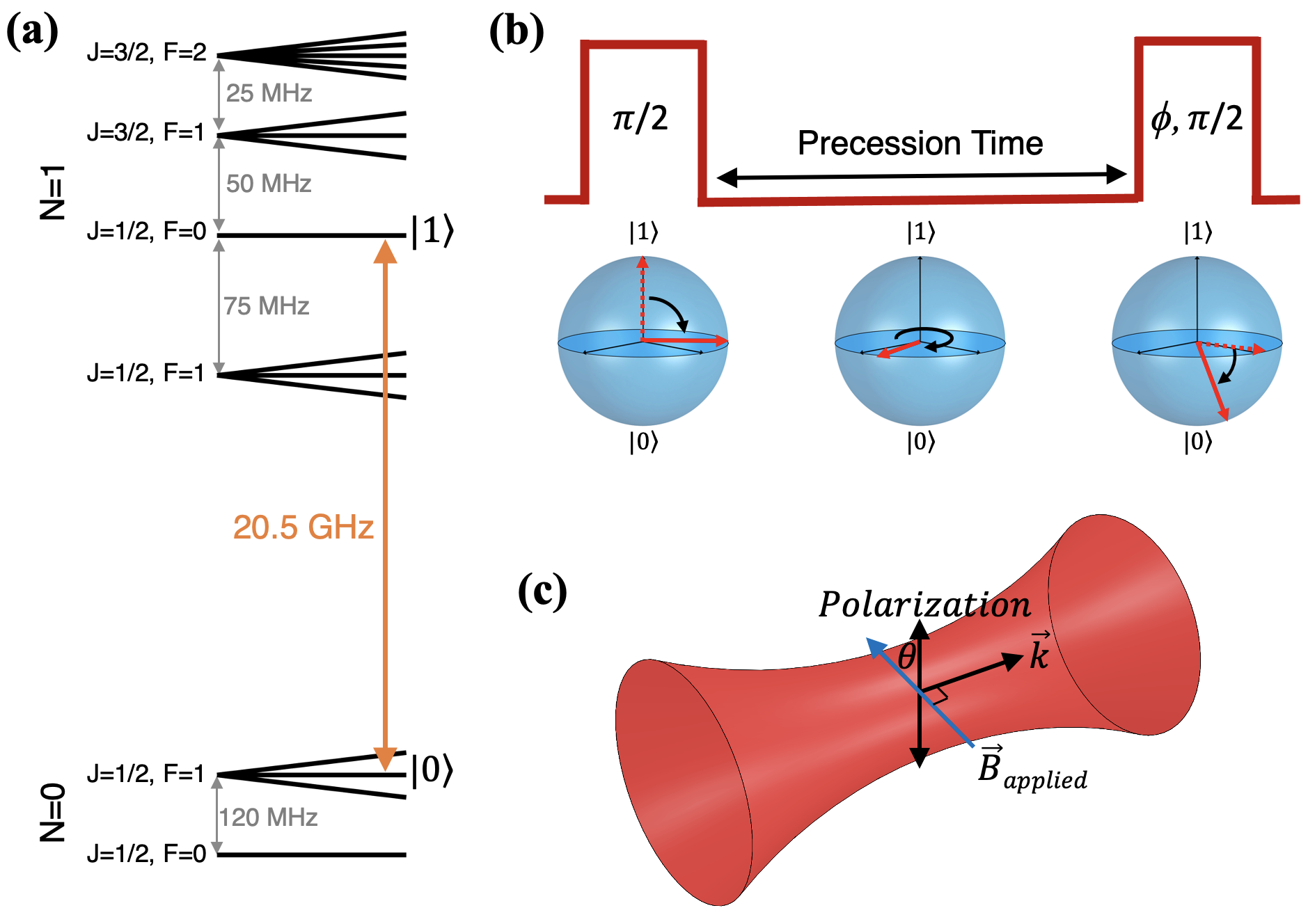}
\caption{\textbf{CaF Structure} (a) The X-state level structure of CaF showing the lower and upper qubit states labeled $|0\rangle$ and $|1\rangle$, respectively. (b) The Ramsey pulse sequence uses a 10~$\mu s$ $\frac{\pi}{2}$-pulse to create a coherent superposition of $|0 \rangle$ and $|1 \rangle$ states. The Bloch vector precesses, accumulating phase until the second $\frac{\pi}{2}$-pulse with a phase shift ($\phi$) rotates the Bloch vector back into the measurement basis prior to imaging the N=1 population. (c) The tweezer light polarization and applied magnetic field lie in the plane perpendicular to the k-vector of the tweezer light.}
\end{figure}

One approach to utilizing molecules as qubits is rearrangeable optical tweezer arrays~\cite{lukin16array, barredo16array, Anderegg2019tweezer, Liu2018}. The dipole-dipole coupling between molecules and associated long-lived excited rotational states have robust coherence properties~\cite{Ni2018,Hughes2020}. For effective 2-qubit gate operations, the rotational coherence time needs to be significantly longer than the gate time. The environment of the molecule can induce decoherence from sources such as fluctuating electric fields, magnetic fields, and inhomogeneous differential light shifts from the optical tweezer light~\cite{Rosenband:18,Kotochigova2010}. To date, all previous studies of molecular coherence times have been done in a bulk gas or lattices~\cite{Bause2020,Blackmore2020,Lin2021,Yedipolecoupling,Seesselberg2018,Neyenhuis2012}. In those works, the coherence times were limited below 10 ms by inhomogeneous broadening from the trapping light or density dependent dipolar scattering. 

In this paper, 
we measure the Ramsey and spin-echo rotational qubit state coherence time of CaF $^2 \Sigma$ molecules in tightly focused 780 nm optical tweezer traps. We mitigate decoherence by identifying and using a set of first-order field insensitive qubit states in CaF. The second order light shift splits the qubit states leading to a light intensity dependent detuning. This is found to limit the coherence time, but is mitigated by fine-tuning the tweezer light polarization and the applied magnetic field to a magic angle. We observe rotational coherence times in an optical tweezer trap of 93(7) ms, approximately 2 orders of magnitude longer than expected 2-qubit gate times for tweezer trapped molecules~\cite{Ni2018,Hughes2020} and over an order of magnitude longer than previously reported rotational coherence times in optical~\cite{Seesselberg2018} and magnetic traps~\cite{Caldwell2020}.

\begin{figure}[b]
\centering
\includegraphics[width=0.475\textwidth]{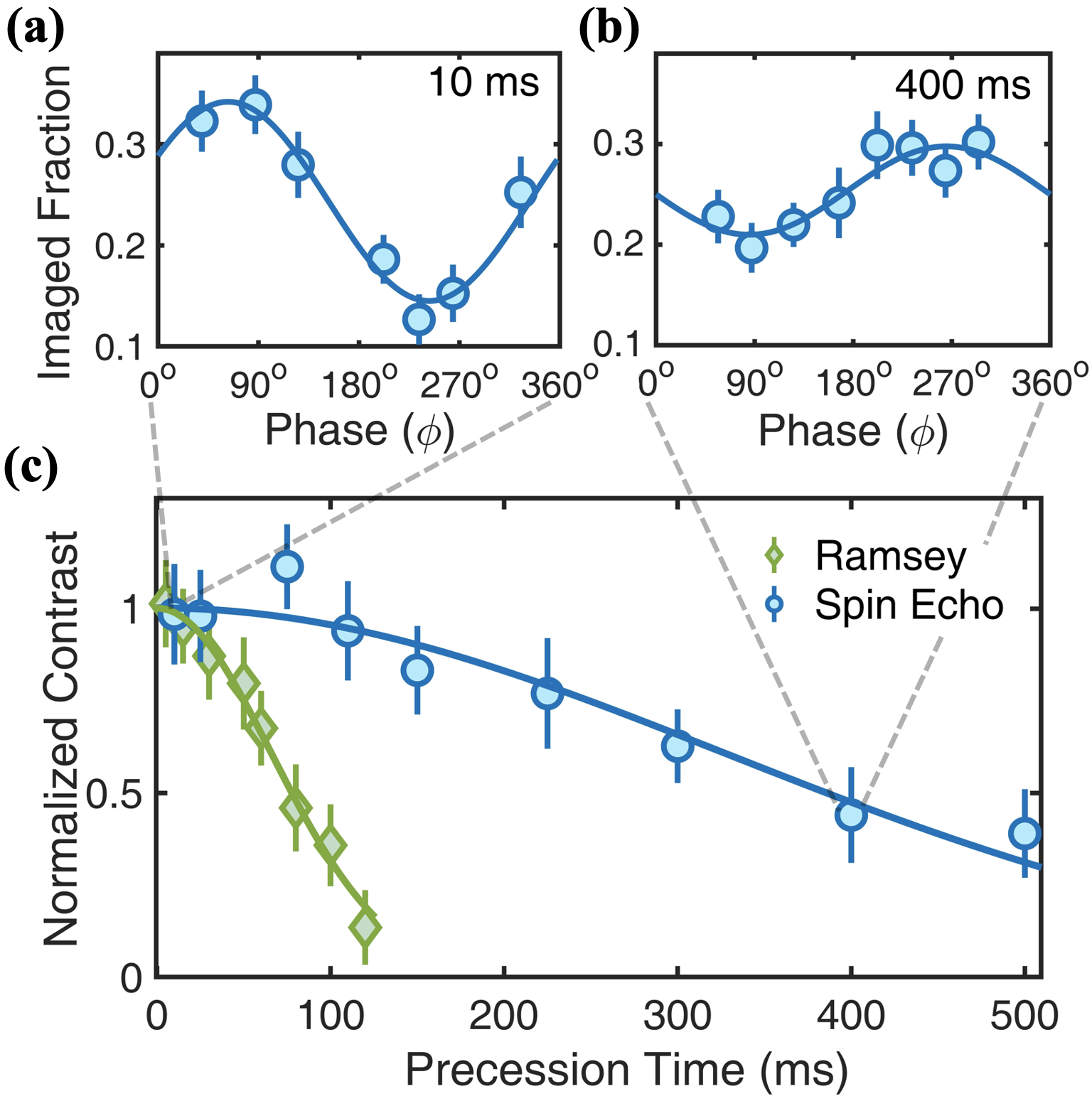}
\caption{\textbf{Coherence Time} (a,b) During the Ramsey and spin-echo pulse sequences, we scan the phase ($\phi$) of the second $\frac{\pi}{2}$-pulse for a fixed precession time. The projection of the wavefunction on $|1 \rangle$ oscillates as a function of $\phi$, which is measured by imaging the population in $|1 \rangle$. (c) The contrast is fit for several precession times. We extract the decay of the contrast using a gaussian $C(t)=e^{-t^2/T^2_c}$ model, producing 1/e-coherence times of $T^{*}_{2}$=93(7) ms and $T_2$=470(40) ms for the Ramsey and spin-echo pulse sequence, respectively. }
\end{figure}

We choose $|0\rangle=|N=0,J=1/2,F=1,m_f=0\rangle$ and  $|1\rangle=|N=1,J=1/2,F=0,m_f=0\rangle$ as the rotational qubit states because they are insensitive to a variety of decoherence mechanisms and offer a dipole moment of approximately 1 Debye. 
To initially prepare the qubits we generate a MOT of $^{40}$Ca$^{19}$F molecules~\cite{anderegg17}, use lambda-enhanced grey molasses cooling~\cite{Cheuk18} to transfer the molecules into a 1064 nm optical dipole trap, and then load the 780 nm tweezers~\cite{Anderegg18} using a secondary lambda-cooling light pulse. Light assisted collisions create a collisional blockade, resulting in single molecule loading as described in ref.~\cite{Anderegg2019tweezer}. The tweezers each contain a single molecule, which are then optically pumped to the $|1 \rangle$ state.  The tweezer depth $U_i$ is then ramped from $U_i=1800 ~\mu$K to $U_f=26~ \mu$K, which results in adiabatic cooling of the molecule from a temperature of $T_i=40 ~\mu$K to $T_f=5~\mu$K, while maintaining $\eta=U/T >5$.

To measure the Ramsey rotational coherence time (T$^{*}_2$), we apply a Ramsey pulse sequence, consisting of two microwave (MW) $\frac{\pi}{2}$-pulses separated by a variable free precession time, shown in figure 1b. The 20.5 GHz resonant MWs are generated by mixing two source, one at 18.5 GHz and the other at 2 GHz.  Both MW sources are synchronized to an external 10 MHz oven-stabilized quartz oscillator reference to ensure stability of the MW phase. Employing a phase shifter on the 2 GHz output, the phase of the second $\frac{\pi}{2}$-pulse can be scanned, functioning as the read out arm of the Ramsey interferometer. After the Ramsey pulse sequence, the molecules are imaged using lambda-imaging~\cite{Cheuk18}. Only the N=1 rotational manifold is near resonant with the imaging light therefore the imaging step measures the projection of the wavefunction on the $|1 \rangle$ state. We scan the phase of the second $\frac{\pi}{2}$-pulse over a full $2 \pi$ cycle at a fixed precession time and then fit the contrast of the resulting sinusoid. To determine the coherence time, we measure the the contrast at several precession times. 

During the free precession time, the detuning between the qubit states can vary due to several different sources of electromagnetic fields and fluctuations. We describe these processes with terms in the Hamiltonian labeled as $\Delta_i (t) \sigma_z$, with $i$ indexing each independent source contributing separately to changes in the relative phase between the MW source and the Bloch vector. Coupling terms proportional to $\sigma_{x,y}$ and $T_1$ population relaxation processes are also possible, however we do not see those effects on the time scales explored in this work.

Changing magnetic fields in the lab environment are one potential source of decoherence. The qubit states have a quadratic Zeeman splitting, which suppresses sensitivity to fluctuations in magnetic field. We apply a field of 1.5 Gauss to split the N=0,F=1 in order to spectroscopically resolve the $|0 \rangle$ state. At 1.5 gauss, the maximum sensitivity to small change in magnetic field is equivalent to a magnetic moment of $\mu_B/10$. The dominant source of magnetic field variation are the fringing fields  from the lab power sources (``line noise''), which have a frequency of 60 Hz. We mitigate the impact of line noise by synchronizing the experimental sequence to the line phase. To address slower magnetic field variations, we implement a fluxgate probe based 3-axis active cancellation around our experimental chamber, suppressing variations in magnetic field well below the 100 $\mu$Gauss level. These steps eliminate magnetic field sources as the main driver of decoherence.

\begin{figure}[b]
\centering
\includegraphics[width=.5\textwidth]{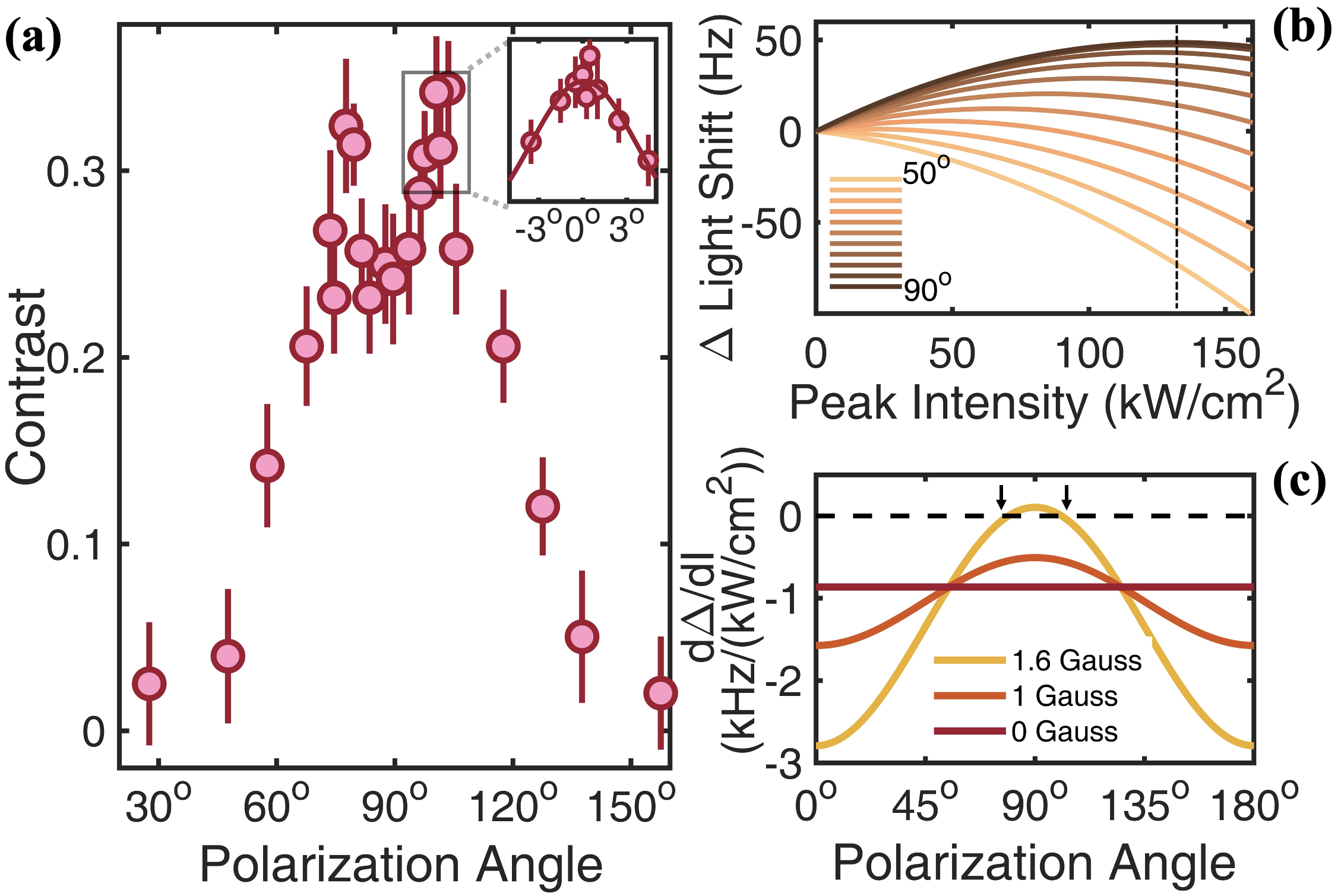}
\caption{\textbf{Magic Angle} (a) We measure the contrast at a fixed precession time of 30 ms and scan the polarization angle between the applied magnetic field and tweezer polarization. Two magic angles appear on either side of $90^\circ$, split by $25(4)^\circ$. The inset shows a fine scan of the magic angle at a precession time of 60 ms. (b) The differential light shift as a function of intensity is initially increasing linearly, but then curves down resulting in a zero-slope region that is first-order insensitive to changes in intensity. Tuning the magic angle moves the zero-slope region near the tweezer intensity of 130 kW/cm$^2$ (vertical line), where the coherence time is maximized. (c) The slope of the light shift (differential polarizability), calculated for a magnetic field of 1.6 gauss, has two zero crossings ($\theta_m$) as a function of angle. Thermal averaging of the light intensity sampled by the molecules in the tweezer shifts and broadens the magic angle, as explained in the text.}
\end{figure}

Confining forces in an optical dipole trap arise from gradients in the AC-electric field of the light. As the molecules move in the tweezer, the light intensity they experience varies. This, when combined with differential polarizability, results in a time varying detuning of the molecule, i.e. changes in the energy spacing between $|0 \rangle$ and $|1 \rangle$.
The differential polarizability arises from the rotational angular momentum of the $|0\rangle$ and $|1 \rangle$ states.  Although all states in the ground rotational manifold have the same scalar polarizability, the states in the excited rotational manifold have a dependence on the projection of angular momentum along the light polarization axis, called tensor polarizability. The $|1 \rangle$ state is spherically symmetric and does not have a first order differential shift. A tensor stark term, however, mixes the $|N=1,J=3/2,F=2,m_f=0 \rangle$ state into $|1\rangle$, leading to a hyperpolarizability ($\beta$), which endows the $|1 \rangle$ state with a quadratic light shift. $\beta$ can be modified significantly by applying a magnetic field directed at an angle $\theta$ to the light polarization. For low light intensities, where the light shift is smaller than the Zeeman shift, the magnetic field is the dominant quantization axis and the corresponding light shift scales linearly with intensity. At high intensities, the light polarization direction becomes the preferred axis, and the hyperpolarizability dominates. When the applied fields are such that the linear polarizability and hyperpolarizability have opposite sign, the light shift has a crossover point, resulting in zero slope with respect to intensity (see figure 3c). Operating in this regime minimizes decoherence effects from the tweezer light, although they remain finite because of the spread in intensities seen by the molecules as they orbit in the trap. To maximize the coherence time in a tweezer with a given intensity, the magnetic field strength and angle can be tuned to the aptly called ``magic angle'' $(\theta_m)$.

In figure 3a, we determine for a fixed tweezer light intensity the polarization angle dependence of the decoherence between the $|0 \rangle$ and $|1\rangle$ states by varying the angle between the magnetic field and the light polarization and then measuring the contrast of the Ramsey fringe after a 30 ms precession time. With a magnetic field of 1.5 Gauss and peak trap light intensity of $130$ kW/cm$^2$, we measure two $\theta_m$'s, split by $25(4)^\circ$. $\theta_m$ can be determined more precisely by measuring the contrast vs. $\theta$ after a longer precession time, shown in figure 3b. At the larger $\theta_m$, we measure the contrast at several precession times and fit the decay of the contrast to a Gaussian model, with a 1/e-coherence time of 93(7) ms. We use Monte-Carlo methods to model decoherence in the trap by sampling an ensemble of single molecules on classical trajectories in a gaussian shaped laser beam. Our model predicts two $\alpha_m$'s, at $77^\circ$ and $103^\circ$, both with a coherence time of 100 ms. The agreement between model and experiment indicates that the coherence time is limited by differential light shifts originating from varying light intensities sampled due to thermal motion in the tweezer, combined with the residual differential polarizability at $\theta_m$.

Figure 4, shows the Ramsey coherence time at temperatures between 40 $\mu$K and 120 $\mu$K. The temperature is varied by adjusting the lambda-cooling light~\cite{Cheuk18,Anderegg2019tweezer}. In this data set, the magic angle is first optimized for the coldest temperature point and remains fixed as we vary the temperature of the molecules. We observe the decoherence rate (reciprocal of the coherence time) increases with temperature. This behavior is attributed to the broadening and shift of the light intensity distribution towards lower intensity, where the differential light shift is no longer flat. Our Monte Carlo simulation, with no free parameters, shows the decoherence rate depends linearly on the temperature, as plotted in figure 4b.
 
Further cooling of the molecule would significantly decrease decoherence. At lower temperatures, where the tweezer is approximately harmonic, the average kinetic energy is equal to potential energy which is proportional to the tweezer light intensity in an optical trap.  For an ideal gas, the variance in energy is T$^2$ times the heat capacity, thus the width of the distribution of intensities experienced by the molecules scales linearly with temperature. Combined with the quadratic nature of the differential light shift at the magic angle, the decoherence rate scales at T$^2$. Thus, further cooling would decrease the spread of the light intensities in this quadratic regime, resulting in a longer coherence time.

To further study the rotational coherence properties of the molecule-tweezer system, we implement a spin-echo (T$_2$) by adding a microwave $\pi$-pulse centered between the two $\frac{\pi}{2}$-pulses. At the same magic angle found for the longest measured Ramsey coherence time, the spin-echo T$_2$ rotational coherence time is 470(40) ms. The varying light shift described before, when averaged over the motion through the trap, reflects the variance in energy of a single molecule in the tweezer and gives rise to shot-to-shot fluctuations of the effective detuning which is suppressed by the spin-echo. We measure similar T$_2$ times for polarization angles as far away as $45^\circ$ degrees from the magic angle, suggesting a different limitation to the spin-echo rotational coherence time.

\begin{figure}[t]
\centering
\includegraphics[width=.45\textwidth]{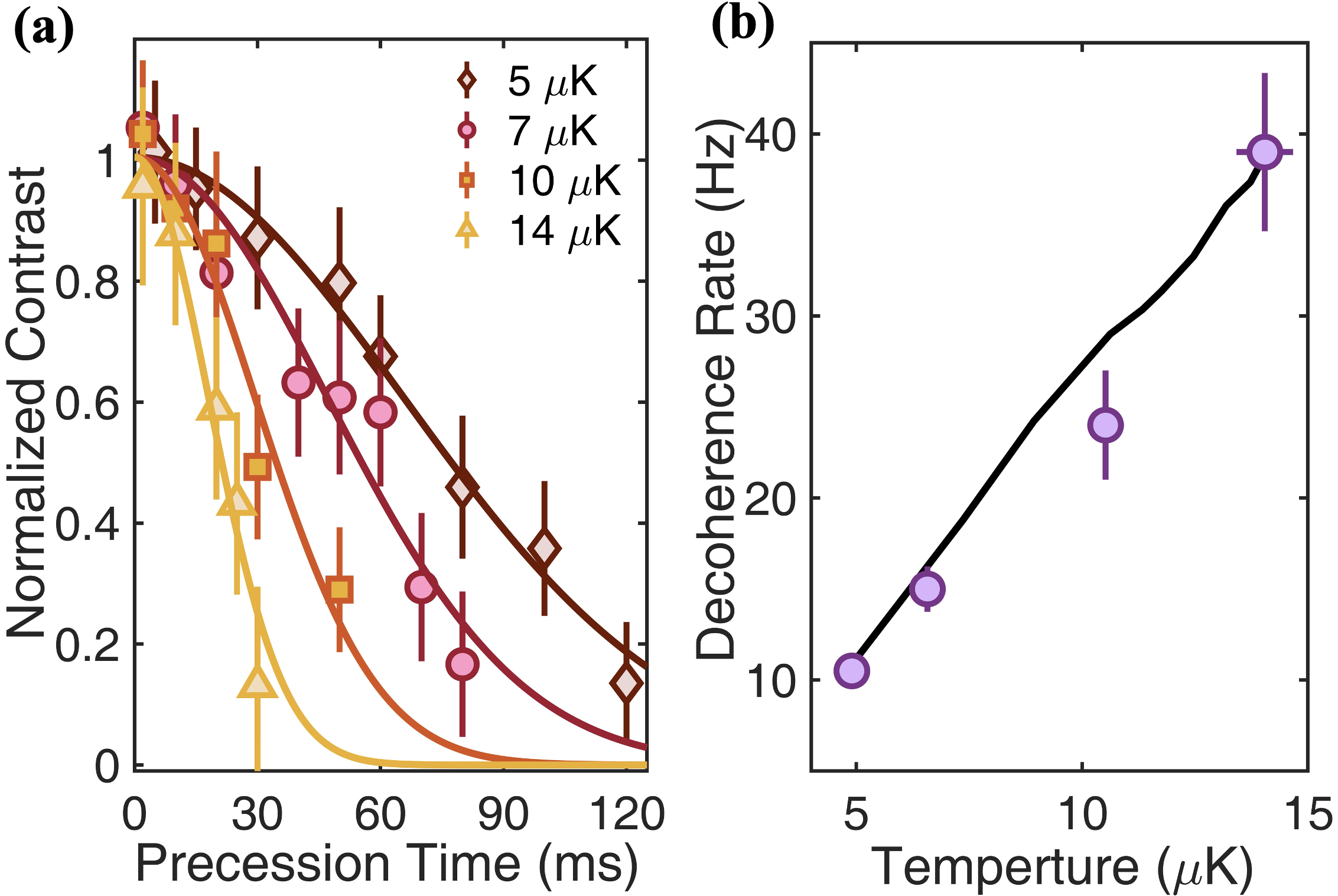}
\caption{\textbf{Temperature Scaling} (a) We measure the decay of contrast for a Ramsey pulse sequence at several different temperatures. The initial temperature is adjusted between 40$\mu$K and 120$\mu$K, then the trap is ramped down to $26~\mu$K resulting in adiabatic cooling of the molecules to the temperatures listed in the plot. (b) The decoherence rate (1/T$^{*}_{2}$) as a function of temperature is plotted. The black line is the result of a Monte Carlo simulation with no free parameters.}
\end{figure}

In conclusion, we demonstrate long rotational coherence times of ultracold polar molecules trapped in optical tweezers. The rotational coherence times reported here are more than an order of magnitude improved over previously measured rotational coherence times and, for the first time, clearly shows a single particle coherence time far exceeding anticipated millisecond-scale dipolar gate times for dipole-dipole coupled molecules in separate optical tweezer traps~\cite{Ni2018,Hughes2020}. Routes towards even faster gate times, e.g. with subwavelength optical tweezers, would allow several thousands of gate operations per coherence time~\cite{Caldwell2020b}. Our implementation of magnetic field cancelation, in combination with the magic angle, leaves residual light shifts from the differential polarizability as the limitation to the Ramsey coherence time on the 100 ms time scale. The future implementation of Raman sideband cooling of the molecules in the tweezers could provide yet another significant improvement to the rotational coherence time~\cite{Caldwell2020a}. The type of qubit states used in our work are generic to $^2 \Sigma$ molecules with nuclear spin $I=1/2$ and MHz scale hyperfine splittings, thus our choice of qubit states is general. Similar approaches could be applied to the many possible laser-coolable polyatomic qubits that have been identified to have similar relevant structure to CaF~\cite{Kozyryev2017,Mitra2020,Yu2019}.

\begin{acknowledgments}
This work was supported by the NSF, DOE, AFOSR, and ARO. SB and SY acknowledge support from the NSF GRFP. LA acknowledges support from the HQI. EC acknowledges support from the NRF of Korea (2021R1C1C1009450, 2020R1A4A1018015)
\end{acknowledgments}

\bibliography{CoherenceTimePaper} 
\end{document}